# On the theory of solitons of fluid pressure and solute density in geologic porous media, with applications to shale, clay and sandstone.

A. Caserta[1], R. Kanivetsky[2], E. Salusti[3]

[1] Istituto Nazionale Geofisica Vulcanologia, Rome, Italy

[2] Department of Bioproducts and Biosystems Engineering, University of Minnesota, 1390 Eckles Ave, St. Paul, MN 55108, USA

[3]INFN Sezione Roma 1, piazzale A. Moro 5, Rome 00185, Italy.

Corresponding author: A. Caserta, Istituto Nazionale Geofisica Vulcanologia, Rome, Italy.

**Abstract.** In this paper we propose the application a new model of transients of pore pressure $p$ and solute density $\rho$ in geologic porous media. This model is rooted in the non-linear waves theory, the focus of which is advection and effect of large pressure jumps on strain (due to large $p$ in a non linear version of the Hooke law). It strictly relates $p$ and $\rho$ evolving under the effect of a strong external stress. As a result, the presence of quick and sharp transients in low permeability rocks is unveiled, i.e. the non-linear Burgers solitons. We therefore propose that the actual transport process in porous rocks for large signals is not the linear diffusion, but could be governed by solitons. A test of an eventual presence of solitons in a rock is here proposed, and then applied to Pierre Shale, Bearpaw Shale, Boom Clay and Oznam-Mugu silt and clay. A quick analysis showing the presence of solitons for nuclear waste disposal and salty water intrusions is also analyzed. Finally, in a kind of "theoretical experiment" we show that solitons could also be present in Jordan and St. Peter sandstones, thus suggesting the occurrence of osmosis in these rocks.

## 1. Introduction

Recent research on geologic porous media has revealed how semi-permeable membranes can create chemical gradients related to osmosis (Marine and Fritz, 1981; Alexander, 1990; House and Pritchett, 1995; Nunn, 1997; Neuzil, 2000; Neuzil and Provost, 2009; Hart, 2012). This process occurs not only in low permeability rocks (e.g., clay, shale and so on), but also in limestone, dolomite and low permeability concrete. To quantify the effect of chemical gradients Ghassemi and Diek (2003, hereafter GD03) developed non-Osanger (Osanger, 1931) analytical models to describe fluid transport forced by osmosis and pore pressure disequilibria around a borehole.

As a follow up Merlani et al. (2011, MSV in the following) applied a non-linear version of the GD03 equations fully considering advection. They demonstrated the presence of quick, large and sharp transients of $p$ and $\rho$, the Burgers solitons (Whitham, 1974) among the solutions of the GD03 model if advection is considered. We here focus attention on such solutions.

The purpose of this study is to discuss the evolution of $p$ and $\rho$ in geologic porous media if solitons are present. To that end we analyze the evolution of $\rho$ and $p$ in clay, shale, and finally sandstones to verify if solitons can occur in these rocks. In various previous studies solitons were invoked to explain the transport of magma in the crust (Scott and Stevenson, 1984; Wiggins and Spiegelman, 1995) and of fluids in sedimentary basins (Connolly and Podladchikov, 1998, 2014). Their presence indeed supports the presence of osmosis in various geologic media, including sandstones. Our model can also enlighten the role of osmosis in the development of anomalous pressures in rocks, not easily explicable in terms of topographic or fluid-density effects (Neuzil, 2000).

In the following we describe the early GD03 and MSV models (Sections 2 and 3) and in Section 4 the case of pressure transients strong enough to increase the matrix strain. A completely novel viewpoint is thus introduced, which considers non-linear advection as well as the effects of a large pore pressure on strain. Applications of our model to Pierre Shale, Bearpaw Shale, Boom Clay and Oxnard-Mugu silt and clay are presented in Sections 5-10. Finally, we use our model in Section 11 to determine the potential occurrence of solitons, i.e., osmosis in sandstones.

## 2. The early models.

The transport processes in porous media were analyzed by Rice and Cleary (1976) with a linear model of isothermal pressure transients. The temperature $T$ was added to the above model by Mc Tigue (1986). In a further development Bonafede (1991) used a similar model for the analysis of waves of $p$ and $T$ in porous rocks. Natale and Salusti (1996) made a first analysis of non-linear effects due to convective transport in these models. The conceptual model considered in this study is essentially that of GD03, the non linear version of MSV.

To investigate the case of one pollutant with density $\rho$ dissolved in the fluid, isothermal processes are here considered since temperature gradients in freshwater geologic systems do not seem to play a critical role. Mc Tigue (1986) obtained our first equation in a 1-D idealization (of interest since the stress in this case is constant)

$$\frac{\partial p}{\partial t}+E\frac{\partial\rho}{\partial t}+F\frac{\partial^2 p}{\partial x^2}+H\frac{\partial^2\rho}{\partial x^2}=0 \tag{1a}$$

his equation is discussed in (A1), (A5), (A6), (A7) and (A10) in the Appendix A. Thus equation (1a) essentially relates variations of $\rho$ and $p$ in isothermal processes. In more detail, if a process has no initial gradients of pressure, or solute, then the system evolution in (1a) is only described by the classical diffusion equation. But in presence of any gradient of $\rho$ and/or $p$ the result is a transient of both these quantities.

Combining equations (A1)-(A4), (A6) and (A7) in turn give the solute mass conservation

$$\frac{\partial \rho}{\partial t} + M \frac{\partial p}{\partial x}\frac{\partial \rho}{\partial x} + N\left(\frac{\partial \rho}{\partial x}\right)^2 + S\frac{\partial^2 \rho}{\partial x^2} + U\frac{\partial^2 p}{\partial x^2} = 0 \qquad (1b)$$

that determines the effect of advection. Greenberg et al. (1973) obtained similar equations but with a more complicate mathematical structure.

We moreover consider the importance of osmosis in a given matrix by considering $E\rho^*/p^*$ and $H\rho^*/Fp^*$ in (1a). As a possible criterion one can check if one of these two ratios is larger, say, than a data uncertainty of 20-30%, as estimated for the parameters equations (1). This finally suggests that osmosis can play an important dynamical role if this criterion

$$E\rho^* / p^* > 0.3 \quad \text{or} \quad H\rho^* / Fp^* > 0.3 \qquad (2)$$

is satisfied. The definitions and estimated values for these geological parameters and for the *E, F,* ....., *U* synthetic quantities are in Tables I and II, where it is also demonstrated the intrinsic relation

$$F/H = M/N \qquad (3)$$

As initial/boundary conditions many of the above studies considered a fluid saturated porous-permeable rock for $x < 0$ (contaminant $\rho_0 + \rho_*$ and pressure $p_0 + p_*$ at $t \approx 0$) as the "source". The transient was assumed to move towards an adjacent homogeneous patch of porous rock ($x > 0$, contaminant $\rho_0$ and pressure $p_0$ at $t \approx 0$).

## 3. The Burgers solitons.

Assuming, as in MSV, the *ansatz* that

$$F\frac{\partial^2 p}{\partial x^2} + H\frac{\partial^2 \rho}{\partial x^2} = -\frac{d}{dt} f(t) \tag{4}$$

from (1a) we obtain that $p + E\rho = f(t)$, i.e. is a function of time only. Such *ansatz* is suggested by a mathematical analysis of Merlani et al. (2001) about symmetry properties of the equations (1) that identified solutions as $F(x^2/t)$, classical for diffusive phenomena, or $G(x - Ut)$ of less interest since related to rigid translations only.

MSV were able to obtain from (1b) a Burgers-like equation

$$\frac{\partial \rho}{\partial t} + A\left(\frac{\partial \rho}{\partial x}\right)^2 - Z\frac{\partial^2 \rho}{\partial x^2} = 0 \tag{5}$$

calling $A = N - EM$ and $Z = UE - S$. Their solution is ruled by a Reynolds number $\boldsymbol{R} = (2A\rho^*)/Z$, and it is a fundamental characterization (Fig 1) to distinguish between linear and non linear solutions (Whitham, 1974; MSV, Caserta et al., 2013). Indeed if $\boldsymbol{R} > 8\text{-}10$ as for a strong initial density, the solution of (5) for $t > t^*$ is (Fig 2)

$$\begin{cases} \rho = \rho_0 + \rho^* & x < 0 \\ \rho = \rho_0 + \dfrac{x^2}{4At} & 0 < x < x_B(t) = \sqrt{4A\rho^* t} \\ \rho = \rho_0 & x > x_B(t) \end{cases} \tag{6}$$

In (6) the time $t^*$ is necessary to avoid mathematical pathologies and can be seen as a small initial delay related to the arrival of a realistic transient (MSV).

The solute density increases till a front at $x_B(t)$ (Appendix B) which velocity $d\, x_B(t) / d\, t$ is initially large but decreases with time. The corresponding $p$ for $0 < x < x_B$ is

$$p = p_0 - \frac{Ex^2}{4At} + \frac{EF - H}{2A} \ln(\frac{t}{t*}) \qquad x < x_B(t) = \sqrt{4A\rho * t} \qquad 0 < t \tag{7}$$

namely the pressure shows the same front, as a strict consequence of (1). The Darcy fluid velocity in turn is (Fig 3)

$$u_D = -\left(\frac{k}{\phi\eta}\right)\frac{\partial p}{\partial x} = \frac{Ek}{\phi\eta}\frac{\partial \rho}{\partial x} \tag{8}$$

where *k*, $\eta$…. are defined in Table 1. In addition the pressure has a component $\frac{EF-H}{2A}\ln\left(\frac{t}{t^*}\right)$ that somehow reminds the formulation of Sorek (1996), is rather small, has no effect on the fluid velocity, is difficult to be related to physical effects and in the following will be disregarded.

It is important to stress how the above *ansatz* (4) is verified in the solutions (6) and (7).

In comparison with the usual diffusion solutions with a front velocity $\approx \sqrt{\pi Z/t}$ , if $\boldsymbol{R}$ >8 -10 in (5) the non linear front velocity is $\approx \sqrt{4A\rho */t}$. Therefore $\boldsymbol{R}$ is the square of the ratio of the non-linear front velocity over the classical diffusion velocity.

All this depends also from the initial density jump $\rho$* and therefore $\boldsymbol{R}$ can be very large for large initial conditions. It therefore plays a fundamental role on practical grounds to reveal if slow classical diffusion or quick solitons related to osmosis can occur in a given rock. It is moreover evident that these *E, F..............., U, A* and *Z* are rather poorly known quantities and therefore they need to be considered critically.

We now test these ideas to other low permeability rocks generalizinig an early study of Caserta et. al. (2013) about Perre Shales.

**4. The effect of a large initial pressure.**

We now discuss thecase of a very large initial pressure: in equations (1) we consider a further pressure term $p \rightarrow \Phi(p) = p + \eta^* p^2$ as a quadratic correction of the linear "Hooke law" (A10). This $\eta^*$ could be very small for elastic or brittle behaviour. But for a ductile behaviour we approximately have from (A10) that $\eta^* = L\ p_y < 0$ where $p_y$ is a parameter rather similar to the pressure yield ultimate fracture strength due to a large initial input (Fig 4) and $L$ a suitable constant.

The equations of the novel model therefore are

$$\begin{cases} E\dfrac{\partial\rho}{\partial t} + \dfrac{\partial\Phi(p)}{\partial t} + F\dfrac{\partial^2 p}{\partial x^2} + H\dfrac{\partial^2 \rho}{\partial x^2} = 0 \\ \dfrac{\partial\rho}{\partial t} + M\dfrac{\partial p}{\partial x}\dfrac{\partial\rho}{\partial x} + N\left(\dfrac{\partial\rho}{\partial x}\right)^2 + S\dfrac{\partial^2\rho}{\partial x^2} + U\dfrac{\partial^2 p}{\partial x^2} = 0 \end{cases} \tag{9}$$

To check if the terms in $S$ and $U$ can be again disregarded in comparison with the non linear terms, we repeat the previous estimate and find that $\boldsymbol{R_{nl}}$, namely the ratio

$$\frac{N(\frac{\partial\rho}{\partial x})^2 - M E\frac{\partial\Phi(\rho)}{\partial x}\frac{\partial\rho}{\partial x}}{S\frac{\partial^2\rho}{\partial x^2} - EU\frac{\partial^2\rho}{\partial x^2}} \approx \frac{N\rho}{S}$$

plays a role similar to $\boldsymbol{R}$ in (5), and thus $S$ and $U$ will be again disregarded if $\boldsymbol{R_{nl}} > 8\text{-}10$.

To solve the system (9) we assume $p = \Gamma(\rho) + X$ , as somehow suggested by (4), and from the second equation of the system (9) for a large $\boldsymbol{R_{nl}}$ we obtain

$$\frac{\partial\rho}{\partial t} + M\frac{\partial[\Gamma(\rho)+X]}{\partial x}\frac{\partial\rho}{\partial x} + N\left(\frac{\partial\rho}{\partial x}\right)^2 = \frac{\partial\rho}{\partial t} + [\ N + \frac{M\,\partial[\Gamma(\rho)+X]}{\partial\rho}]\,(\frac{\partial\rho}{\partial x})^2 \approx 0 \tag{10}$$

Multiplying (10) for $N + \frac{M\,\partial[\Gamma(\rho)+X]}{\partial\rho}$ we thus obtain (MSV) the solution

$$N\rho + M[\Gamma(\rho)+X] = N\rho + M\,p = \frac{x^2}{4\,t} + \text{cos}t \tag{11}$$

Replacing this (11) in (9) from (3) we obtain that

$$-\frac{EM}{N}\frac{\partial p}{\partial t}+\frac{\partial \Phi(p)}{\partial t}+\frac{E\,x^2}{4N\,t^2}+F\frac{\partial^2 p}{\partial\, x^2}+H\frac{\partial^2 \rho}{\partial\, x^2}= \tag{12}$$

$$-\frac{EM}{N}\frac{\partial p}{\partial t}+\frac{\partial \Phi(p)}{\partial t}+\frac{E\,x^2}{4N\,t^2}+\frac{2H}{Nt}=0$$

This (12) is an unexpected relation since in this way the cumbersome system (9) becomes just an algebraic equation. Thus we find

$$p^2 + A\, p/\eta^* N + Ex^2/4\eta^* N\, t = (2H/\, N\eta^*)\, ln\, (t/t^*) \tag{13}$$

and, disregarding again the small term with $(2H\,/\,N\,\eta^*)ln(t/t^*)$, we obtain

$$p \approx -\frac{E\,x^2}{4\,A\,t}-\frac{\eta^* E^2 N}{16\,A}\frac{x^4}{t^2}+\ldots\ldots \tag{14}$$

The first order correction for a large pressure input therefore is a pressure increase $-\frac{E^2N}{16A}\eta^*\frac{x^4}{t^2}$ since $\eta^* < 0$. We moreover remark how often in these rocks $N \approx A \approx 10^{-7}$ while $E \approx -10^5$ in SI and therefore we approximately obtain

$$p \approx -\frac{E\,x^2}{4\,A\,t}-\frac{\eta^* E^2}{16}\frac{x^4}{t^2}+\ldots\approx 10^{12}\frac{x^2}{t}-10^8\eta^*\frac{x^4}{t^2}\ldots\ldots\ldots \tag{15}$$

This stresses the importance of $\eta^*$ also for short times.

## 5. Pierre Shale

The Pierre Shale is a low permeability formation of Upper Cretaceous age. This dark-gray shale is fossiliferous, has maximum thickness of about 210 m and overlies sandstone aquifer systems (Bredehoeft, 1983). The Pierre Shale is correlated with other marine shales that occur farther west, such as the Bearpaw Shale in United States and Canada (Caserta et al., 2013).

The data discussed in Table $I_a$ were obtained in central South Dakota, USA (Barbour and Fredland, 1989; Neuzil, 2000; Simm, 2007; Neuzil and Provost, 2009; Sarout and Detournay, 2011). Its mineralogy is 70-80% of clay, of which about 80% is a mixed-layer of smectite-illite. The shale, at this site, is saturated at the depth of ~ 75 m.

In all these examples the fluid viscosity μ is $3 \times 10^{-4}$ and the fluid bulk modulus $K_f$ is $2 \times 10^{9}$ in SI.

**Table $I_a$.** Estimated values of relevant parameters of Pierre Shale, in SI.

| Parameters | Values in SI | Units |
|---|---|---|
| $\varphi$ Rock porosity | 0.3 | / |
| $k$ Intrinsic permeability | $10^{-18}$ | $m^2$ |
| Θ Solute reflection coefficient | 0.25 | / |
| $D$ Solute diffusion coefficient | $10^{-8}$ | $m^2/s$ |
| $M^s$ Solute molar mass (Na Cl) | 0.06 | kg/mol |
| α Biot coefficient | 0.7 | / |
| $\omega_0$ Swelling coefficient | $10^5$ | Pa |
| $K$ Bulk modulus | $4 \times 10^6$ | Pa |
| $K_s$ Bulk modulus of the solid matrix | $10^7$ | Pa |
| $\bar{\rho}$ Estimated solute density | 1 | $kg/m^3$ |
| $\bar{p}$ Estimated pore pressure | $10^5$ | Pa |

To estimate the swelling coefficient $\omega_0$ we compare the Sarout and Detournay (2011) with GD03 articles and find that the osmotic pressure $\pi(x,t)$ is indeed related to $\omega_0$ as

$$\pi(x,t) \approx \omega_0 \left( \frac{1}{\bar{\rho}} - \frac{1}{\bar{\rho}_D} \right) \rho \qquad (16)$$

and for the shales following GD03 we obtain as an average value $\omega_0 \approx 10^5$ Pa.

**Table I$_b$.** Coefficients of equation (6) for Pierre Shale, in SI.

| Coefficient | Numerical value in SI |
|---|---|
| | |
| $E = -\frac{\omega_0}{\alpha^2 + KV} \left( \frac{1}{\bar{\rho}_S} - \frac{1}{\bar{\rho}_D} \right)$ | $-1.5\text{x}10^5\ m^2/s$ |
| $F = -\left( \frac{k}{\eta} K + \frac{k\Theta K \bar{\rho}_f}{\eta \bar{\rho}_D} \right) / (\alpha^2 + VK)$ | $-2,5\text{x}10^{-8} m^2/s$ |
| $H = \frac{k\ \Theta \bar{\rho}_f R^* TK}{\eta M^S (\alpha^2 + KV)} \left( \frac{1}{\bar{\rho}_D} + \frac{1}{\bar{\rho}_S} \right)$ | $0.2 m^4/s$ |
| $M = -\frac{k}{\phi \eta} \left[ \frac{1}{\bar{\rho}_f} + \frac{\Theta}{\bar{\rho}_D} \right] \bar{\rho}_f$ | $-10^{-14} m^2/s$ Pa |
| $N = \frac{k}{\eta \phi} \Theta \frac{R^* T}{M^S} \left[ \frac{1}{\bar{\rho}_D} + \frac{1}{\bar{\rho}_S} \right] \bar{\rho}_f$ | $10^{-7}\ m^4/s^3$ Pa |
| $S = -\frac{D}{\phi}$ | $-3\text{x}10^{-8} m^2/s$ |
| $U = \frac{M^S D \bar{\rho}_S}{R^* T \phi\ \ \bar{\rho}_f}$ | $7.5\text{x}10^{-16}$ s |
| $V = \frac{\alpha - \phi}{K_s} + \frac{\phi}{K_f} - \frac{\omega_0}{K} \frac{M^S}{R^* T \bar{\rho}_D}$ | $2.5\text{x}10^{-8} Pa^{-1}$ |

The above data show that for Pierre shales $A = N - M\,E \approx 10^{-7}$and $Z = E\,U - S \approx 3x10^{-8}$ in the SI. This gives a small $\boldsymbol{R} \approx 3\rho*$. Therefore solitons can be present if initially $\rho* > 3$ in SI, or equivalently the initial pressure $p* > 10^6$. This also gives from (2) that osmosis plays a role if

$$E\,\rho*\,/\,p* \approx 10^{\,5}\rho*/p* > 0.3 \qquad \text{or} \qquad H\,\rho*/F\,p* \approx 10^{\,9}\rho*/p* > 0.3$$

namely, if $\rho*/p* > 10^{-5}$ in SI. Finally a strong initial pressure implies a positive pressure variation

$$-\frac{\eta * E^2 N}{16\;A}\;\frac{x^4}{t^2} \approx -\;10^8\,\eta*\frac{x^4}{t^2}\,.$$

## 6. Bearpaw Shale

The Bearpaw Formation, also called the Bearpaw Shale, is a sedimentary rock found in Northwest Saskatchewan, east of the Rocky Mountains. To the east and south it blends into the Pierre Shale. The data reported here were measured using samples collected near Saskatoon, Saskatchewan, Canada (Cey et al, 2001). This site consists of a ~ 76 m thick, massive and plastic marine clay (about 5% sand, 38% silt and 57% clay) deposited approximately 70 million years before present, at a depth of about 88 to 123 m.

The mineralogy of the Bearpaw Shale is claystone, somehow similar to Pierre Shale but with total clay of ~57%, of which 50-60% is smectite with lesser amount of illite (10-20%). Indeed Pierre Sale and this Bearpaw Shale are somehow similar rocks from a mineralogy and origin point of view (Barbour and Fredland, 1989; Cey et al., 2001; Simm, 2007; Neuzil and Provost, 2009). Therefore the values are tentatively assumed to be as those of Pierre Shale.

**Table II$_a$.** Estimated values of relevant parameters for Bearpaw Shale.

| Parameters | Values in SI | Units |
|---|---|---|
| $\varphi$ Rock porosity | 0.4 | / |
| $k$ Intrinsic permeability | $5\text{x}10^{-21}$ | $m^2$ |
| $\Theta$ Solute reflection coefficient | 0.25 | / |
| $D$ Solute diffusion coefficient | $5\text{x}10^{-10}$ | $m^2/s$ |
| $M^s$ Solute molar mass (Na Cl) | 0.06 | kg/mol |
| $\alpha$ Biot coefficient | 0.6 | / |
| $\omega_0$ Swelling coefficient | $2\text{x } 10^6$ | Pa |
| $K$ Bulk modulus | $4\text{x}10^6$ | Pa |
| $Ks$ Bulk modulus of the solid matrix | $1.5\text{x}10^7$ | Pa |
| $\bar{\rho}$ Estimated solute density | 16.0 / 33.0 | $kg/m^3$ |
| $\bar{p}$ Estimated pore pressure | $2.4\text{x}10^5$ | Pa |

From these estimates we assume an average $\bar{\rho} \approx$ 25kg/m$^3$.

**Table II$_b$.** Coefficients of equations (1) for Bearpaw Shale, in SI

| Coefficient | Numerical value in SI |
|---|---|
| $E = -\dfrac{\omega_0}{\alpha^2 + KV}\left(\dfrac{1}{\bar{\rho}_S} - \dfrac{1}{\bar{\rho}_D}\right)$ | $-3\text{x}10^5\ m^2/s$ |
| $F = -\left(\dfrac{k}{\eta}K + \dfrac{k\Theta K\bar{\rho}_f}{\eta\bar{\rho}_D}\right) / (\alpha^2 + VK)$ | $-2\text{x}10^{-10}\ m^2/s$ |
| $H = \frac{k\Theta\bar{\rho}_f R^*TK}{\eta M^S(\alpha^2+KV)}(\frac{1}{\bar{\rho}_D}+\frac{1}{\bar{\rho}_S})$ | $7x10^{-5}$ $m^4/s$ |
| $M = -\frac{k}{\phi\eta}[\frac{1}{\bar{\rho}_f}+\frac{\Theta}{\bar{\rho}_D}]\bar{\rho}_f$ | $-5x10^{-17}$ $m^2/s$ Pa |
| $N = \frac{k}{\eta\phi}\Theta\frac{R^*T}{M^S}[\frac{1}{\bar{\rho}_D}+\frac{1}{\bar{\rho}_S}]\bar{\rho}_f$ | $10^{-11} m^4/s^3$ Pa |
| $S = -\frac{D}{\phi}$ | $-10^{-9}$ $m^2/s$ |
| $U = \frac{M^S D\bar{\rho}_S}{R^*T\phi\ \bar{\rho}_f}$ | $7.5x10^{-16}$ s |
| $V = \frac{\alpha-\phi}{K_s}+\frac{\phi}{K_f}-\frac{\omega_0}{K}\frac{M^S}{R^*T\bar{\rho}_D}$ | $1.5x10^{-8}$ $Pa^{-1}$ |

Thus for $\bar{\rho}$ ~ 25 we have $A = N - M E \approx 10^{-11}$and $Z = E U - S$ $10^{-9}$ and $\boldsymbol{R} \approx 0{,}03$.

To obtain in the Bearpaw Shale a soliton is thus necessary a very large value $\rho^* \approx 300$, much larger than that of Pierre Shale. All this implies that if a waste disposal facility placed in Bearpaw Shale would be in contact with some external fluid, a larger impact of pressure in comparison with those for the Pierre Shale is necessary to generate such large and quick transients.

This also gives from the criterion (2) that osmosis plays a role if

$E\,\rho^*/p^* \approx 10^5\rho^*/p^* > 0.3$ or $H\,\rho^*/F\,p^* \approx 10^6\rho^*/p^* > 0.3$ namely if $\rho^*/p^* > 10^{-5}$in SI.

For a strong initial pressure one has again a positive jump

$$-\frac{\eta^* E^2 N}{16\ A}\ \frac{x^4}{t^2} \approx -\ 10^8\,\eta^*\frac{x^4}{t^2}\ .$$

## 7. Boom Clay

The Oligocene Boom Clay is a silty clay of marine origin occurring in north-eastern Belgium. The Boom Clay is a plastic clay (19–26 %) with total porosity of 0.35–0.40. The average clay mineral content (phyllo-silicates) is 60% water in weight and the clay mineralogy is dominated by illite/smectite mixed layers with a detail mineralogy of clay 30-70%, smectite 10-30%, mixed layer smectite-illite 5-50%, illite 10-30%, and chlorite 1-5% (Helgerud et al., 1999; Garavito et al., 2007; Neuzil and Provost, 2009; Delage et al., 2010) .

**Table III$_{a.}$** Estimated values for relevant parameters of Boom Clay, in SI.

| Parameters | Values in SI | Units |
|---|---|---|
| $\varphi$ Rock porosity | 0.4 | / |
| $k$ Intrinsic permeability | $5x10^{-18}$ | $m^2$ |
| $\Theta$ Solute reflection coefficient | 0.3 | / |
| $D$ Solute diffusion coefficient | $4x10^{-6}$ | $m^2/s$ |
| $M^s$ Solute molar mass (Na Cl) | 0.06 | kg/mol |
| $\alpha$ Biot coefficient | 0.6 | / |
| $\omega_0$ Swelling coefficient | $5x10^5$ | Pa |
| $K$ Bulk modulus | $8 \times 10^6$ | Pa |
| $Ks$ Bulk modulus of the matrix | $2 x10^7$ | Pa |
| $\bar{\rho}$ Estimated solute density | 5 | $kg/m^3$ |
| $\bar{p}$ Estimated pore pressure | $2x10^5$ | Pa |

We therefore have

**Table III$_b$.** Coefficients of equations (1) for Boom Clay, in SI.

| Coefficient | Numerical value in SI |
|---|---|
| $E = -\frac{\omega_0}{\alpha^2 + KV}\left(\frac{1}{\bar{\rho}_S} - \frac{1}{\bar{\rho}_D}\right)$ | $-2\times10^{5}$ m$^2$/s |
| $F = -\left(\frac{k}{\eta}K + \frac{k\Theta K\bar{\rho}_f}{\eta\bar{\rho}_D}\right) / (\alpha^2 + VK)$ | $-3\times10^{-7}$ m$^2$/s |
| $H = \frac{k\Theta\bar{\rho}_f R^*TK}{\eta M^S(\alpha^2 + KV)}\left(\frac{1}{\bar{\rho}_D} + \frac{1}{\bar{\rho}_S}\right)$ | 0.8 m$^4$/s |
| $M = -\frac{k}{\phi\,\eta}\left[\frac{1}{\bar{\rho}_f} + \frac{\Theta}{\bar{\rho}_D}\right]\bar{\rho}_f$ | $-5\times10^{-14}$ m$^2$/s Pa |
| $N = \frac{k}{\eta\phi}\Theta\frac{R^*T}{M^S}\left[\frac{1}{\bar{\rho}_D} + \frac{1}{\bar{\rho}_S}\right]\bar{\rho}_f$ | $10^{-7}$m$^4$/s$^3$ Pa |
| $S = -\frac{D}{\phi}$ | $-10^{-5}$ m$^2$/s |
| $U = \frac{M^S D\bar{\rho}_S}{R^*T\phi\ \bar{\rho}_f}$ | $10^{-12}$ s |
| $V = \frac{\alpha - \varphi}{K_s} + \frac{\varphi}{K_f} - \frac{\omega_0}{K}\ \frac{M^S}{R^*T\bar{\rho}_D}$ | $10^{-8}$ Pa$^{-1}$ |

From these values we obtain $A = N - EM \approx 10^{-7}$and $Z = EU - S \approx 10^{-5}$ and therefore $\boldsymbol{R}$ is about $0.01\rho^*$. Also in this case shock solitary waves are possible, but for a very large initial stress $\rho^* \approx$ $10^3$ or an external pressure $p^*$ about $10^7$ in SI. Thus osmosis plays a role for $E\,\rho^*/p^* \approx 10^5\rho^*/p^* >$ 0.3 or $H\,\rho^*/F\,p^* \approx 10^7\rho^*/p^* > 0.3$ , consequently osmosis is important if $\rho^*/p^* > 10^{-7}$ .

One can therefore have Burgers solitons, but for very large values of $\rho^*$. All this implies that if a radioactive disposal facility placed in Boom Shale would be in contact with some external fluid, in comparison with those for the Pierre Shale a really larger impact of radioactive pollutants would be necessary to generate such transients (Henrion et al., 1990).

Again a strong initial pressure implies a positive variation very similar to those already found for Pierre shale $-\dfrac{\eta^* E^2 N}{16\ A}\ \dfrac{x^4}{t^2} \approx -\ 10^8\,\eta^*\dfrac{x^4}{t^2}$ .

## 8. Underground waste isolation in the Boom Clay

Clay rich deposits are usually considered to be good natural media for underground waste isolation because of their low permeability (Garavito et al., 2007). In the absence of water conductive features, these deposits provide the environment required for a fully reliable waste containment. Indeed in these deposits the diffusion seems to be the dominant transport process because clay minerals retard the movement of contaminants by ion exchange, sorption and ultra-filtration (De Cannière et al., 1996; Cey et al., 2001). The clay deposits in the total absence of fluid conductive features are therefore key barrier for ensuring the long-term safety of a disposal system. In more realistic situations where some amount of water is present, the comprehensive understanding of the physical and chemical processes controlling the water and solute transport through low-permeability clay type formations is a key step for assessing their suitability as host rocks (Garavito et al., 2007).

Example of such host rock is an over-consolidated marine Oligocene deposit, the Boom Clay (Garavito et al., 2007). In addition, it is often considered as a reference host formation for waste disposal in Belgium because of its favorable characteristics. In fact, for more than 25 years extensive hydraulic, geo-mechanical and geo-chemical research has been carried out on the Boom Clay at the HADES Underground Research Laboratory (URL) in Mol (Belgium). Primary objectives of these experiments have been to characterize the in situ hydrogeological conditions, to determine the hydraulic parameters and to study the mechanisms controlling the chemistry and the composition of the Boom Clay pore water (Baeyens et al., 1985; Henrion et al., 1990; De Cannière et al., 1996).

In situ data however confirm the occurrence of chemical osmosis in low-permeability plastic formations, such as those present in the Boom Clay. The osmotic efficiency of Boom Clay is high under undisturbed chemical conditions but rapidly decreases when the dissolved salts concentration increases. The semi-permeable membrane behavior of the high efficiencies Boom Clay is actually considered to be most important for waste disposal.

In such complex situation our model is directly applicable to waste isolation in underground controlled facilities. The rocks under consideration for repositories must indeed have extremely low permeability and to avoid complex thermal effects the convective heat transfer must be minimal. If the temperature changes at short distance away from a heat-producing waste canister are small, the assumption of constant material properties may be considered as appropriate.

We now test the flexibility of our model with a theoretical simulation, i.e. we compute the variations of Table IIIa and IIIb as if the kind of solute in Boom Clay is not the usual Na Cl but a material with a larger solute molar mass, say, $M^{s} = 0.18$ to simulate dangerous nuclear waste as Cesium 137 Clorure. This could simulate the eventual propagation of rather heavy, soluble material when in contact with the deposit boundary. We also tentatively assume the same increase of about 300% for the values of the solute density and solute reflection coefficient, following the assumption that such coefficients may follow the solute molar mass increase while the permeability *k* and the diffusion *D* are assumed to be inversely dependent for the same 300% estimated value.

**Table IV$_a$** Estimated values for relevant parameters for Boom Clay for a dense solute with a solute molar mass $M^s$= 0.18 in SI.

| Coefficients | Values in SI | Units |
|---|---|---|
| $\varphi$ Rock porosity | 0.4 | / |
| *k* Intrinsic permeability | $5\times10^{-18}\rightarrow 2\text{x}10^{-18}$ | $m^2$ |
| Θ Solute reflection coefficient | 0.3→ 1 | / |
| *D* Solute diffusion coefficient | $4 \text{ x } 10^{-6}\rightarrow 10^{-6}$ | $m^2/s$ |
| $M^s$ Solute molar mass (Na Cl) | 0.06 →0.18 | kg/mol |
| $\alpha$ Biot coefficient | 0.6 | / |
| $\omega_0$ Swelling coefficient | $10^5$ | Pa |
| *K* Bulk modulus | $8 \text{ x } 10^6$ | Pa |
| *Ks* Bulk modulus of the matrix | $2 \text{ x } 10^7$ | Pa |
| $\bar{\rho}$ Estimated solute density | 5→ 15 | $Kg/m^3$ |
| $\bar{p}$ Estimated pore pressure | $2\text{x}10^5$ | Pa |

In this kind of numerical simulation we have

**Table IV$_{b.}$** Coefficients of equation (1) for Boom Shale, in SI.

| Coefficient | Numerical value in SI |
|---|---|
| $E = -\frac{\omega_0}{\alpha^2 + KV}\left(\frac{1}{\overline{\rho}_S} - \frac{1}{\overline{\rho}_D}\right)$ | $- 2 \times 10^4$ m²/s |
| $F = -\left(\frac{k}{\eta}K + \frac{k\Theta K\overline{\rho}_f}{\eta\overline{\rho}_D}\right)\ /(\alpha^2+VK)$ | $-2\times10^{-7}$ m²/s |
| $H = \frac{k\Theta\overline{\rho}_f R^*TK}{\eta M^S}\left(\frac{1}{\overline{\rho}_D} + \frac{1}{\overline{\rho}_S}\right) / (\alpha^2+KV)$ | 0.1 m⁴/s |
| $M = -\frac{k}{\phi\eta}\left[\frac{1}{\overline{\rho}_f} + \frac{\Theta}{\overline{\rho}_D}\right]\overline{\rho}_f$ | $- 3 \times 10^{-14}$ m²/s Pa |
| $N = \frac{k}{\eta\phi}\Theta\frac{R^*T}{M^S}\left[\frac{1}{\overline{\rho}_D} + \frac{1}{\overline{\rho}_S}\right]\overline{\rho}_f$ | $2\times10^{-8}$ m⁴/s³ Pa |
| $S = -\frac{D}{\phi}$ | $- 2\times10^{-6}$ m²/s |
| $U = \frac{M^S D\overline{\rho}_S}{R^*T\phi\ \ \overline{\rho}_f}$ | $3\times10^{-12}$ s |
| $V = \frac{\alpha - \phi}{K_s} + \frac{\phi}{K_f} - \frac{\omega_0}{K}\frac{M^S}{R^*T\,\overline{\rho}_D}$ | $10^{-8}$ $Pa^{-1}$ |

In summary, with these heuristic estimates we find that $A = N - EM \approx 2 \text{ x}10^{-8}$ and $Z = EU - S$ is about $2\text{x}10^{-6}$ and, therefore, $A/Z$ is as small as ≈ 0.02 and also ***R*** is a small ≈ 0.02 $\rho^*$. Also in this case, shock solitary waves are possible for this model, but for a very large initial stress $\rho^* \approx 5\text{x}10^2$ or a external pressure stress $p^* \approx 10^8$. This confirms the wise choice of this kind of rock for deep nuclear deposit.

## 9. Oxnard-Mugu Silt and Clay

The late Pleistocene Oxnard-Mugu silt and clay deposits occur in Oxnard coastal basin, Ventura County (California). A confining bed, consisting of silt and clay, separates the Oxnard and Mugu aquifers. These sediments have maximum thickness of 46 m, are well stratified and the strata are composed mainly of silty material. The lateral extent of these rocks is approximately 20-30 km (California Department of Water Resources, 1971).

Laboratory investigation of the pore fluids in such silt and clay strata **(**Greenberg et al.,1973**)** indicated Na Cl concentrations of 1.34-1.92 and in our analysis we assume the average $\rho \approx 1.6$. Coupled salt and water flows have indeed been observed through this low permeability material (i.e. the confining bed, consisting of silt and clay), but no detail mineralogy is studied (Delage et al., 2010; Helgerud et al., 1999).

**Table $V_a$.** Estimated values of relevant parameters for Oxnard-Mugu Silt and Clay, in SI.

| Coefficients | Values in SI | Units |
|---|---|---|
| $\varphi$ Rock porosity | 0.36 | / |
| $k$ Intrinsic permeability | $10^{-16}$ | $m^2$ |
| $\Theta$ Solute reflection coefficient | 0.15 | / |
| $D$ Solute diffusion coefficient | $10^{-8}$ | $m^2/s$ |
| $M^s$ Solute molar mass (Na Cl) | 0.06 | kg/mol |
| $\alpha$ Biot coefficient | 0.45 /0.63 | / |
| $\omega_0$ Swelling coefficient | $2x10^5$ | Pa |
| $K$ Bulk modulus | $10^{-7}$ | Pa |
| $Ks$ Bulk modulus of the matrix | $2x10^{-7}$ | Pa |
| $\bar{\rho}$ Estimated solute density | 1.34 – 1.92 | $Kg/m^3$ |
| $\bar{p}$ Estimated pore pressure | $3x10^5$ | Pa |

Since no detailed mineralogy for this basin is available, we heuristically assume for $\alpha$ and $\Theta$ an average of the values provided for shales and clays here examined.

**Table $V_b$.** Coefficients of equations (1) for Oxnard-Mugu Silt and Clay.

| Coefficient | Numerical value in SI |
|---|---|
| $E=-\frac{\omega_0}{\alpha^2+KV}(\frac{1}{\bar{\rho}_S}-\frac{1}{\bar{\rho}_D})$ | $-2x10^5\ m^2/s$ |
| $F = -(\frac{k}{\eta}K + \frac{k\Theta K\bar{\rho}_f}{\eta\bar{\rho}_D})\ /(\alpha^2+VK)$ | $-10^{-5}$ m$^2$/s |
| $H = \frac{k\Theta\bar{\rho}_f R^*TK}{\eta M^S(\alpha^2+KV)}(\frac{1}{\bar{\rho}_D}+\frac{1}{\bar{\rho}_S})$ | 4 m$^4$/s |
| $M = -\frac{k}{\phi\eta}[\frac{1}{\bar{\rho}_f}+\frac{\Theta}{\bar{\rho}_D}]\bar{\rho}_f$ | $-10^{-12}$ m$^2$/s Pa |
| $N = \frac{k}{\eta\phi}\Theta\frac{R^*T}{M^S}[\frac{1}{\bar{\rho}_D}+\frac{1}{\bar{\rho}_S}]\bar{\rho}_f$ | $10^{-6}$m$^4$/s$^3$ Pa |
| $S = -\frac{D}{\phi}$ | $-3\text{x}10^{-8}$ m$^2$/s |
| $U = \frac{M^S D\bar{\rho}_S}{R^*T\phi\ \bar{\rho}_f}$ | $7\text{x}10^{-16}$ s |
| $V = \frac{\alpha-\phi}{K_s}+\frac{\phi}{K_f}-\frac{\omega_0}{K}\frac{M^S}{R^*T\bar{\rho}_D}$ | $7\text{x}10^{-9}$Pa$^{-1}$ |

The values of Table V$_b$ give $A = N - M\ E \approx 10^{-6}$ and $Z = S - EU \approx 3\text{x}10^{-8}$ and therefore the corresponding $\boldsymbol{R} \approx 10^2$. All this shows that one can easily have cases of solitary waves also for small values of $\rho^*$. In this matrix the osmosis can therefore play a role if

$E\rho^*/p^* \approx 10^5\rho^*/p^* > 0.3$ or $H\rho^*/F\,p^* \approx 10^6\rho^*/p^* > 0.3$.

Therefore osmosis is important if $p^* \approx 10^5$ and this confirms how solitons can be relatively frequent in such systems.

A strong initial pressure again gives a similar positive variation $-\frac{\eta^* E^2 N}{16\ A}\ \frac{x^4}{t^2} \approx -\ 10^8\,\eta^*\frac{x^4}{t^2}$.

## 10. Sea water intrusion in the Oxnard coastal basin

The groundwater regime in the Oxnard coastal basin is a multiple aquifer system of successive confining beds and aquifers, described by Greenberg et al. (1973). The Oxnard aquifer, which is at a depth of about 49 meters and was the principal producing aquifer in the basin, has been intruded since years 1930 by seawater as a result of a general lowering of groundwater levels. This seawater intrusion, which now extends several miles inland, poses a serious threat to the water resources of the region since the Oxnard aquifer has been the principal source of water for the Oxnard area.

Where water from the Oxnard can no longer be used because of the seawater intrusion, wells have been drilled to the deeper Mugu aquifer, which is separated from the Oxnard by an aquitard layer of fine-grained material. Possible consequences of the seawater diffusion as discussed by Greenberg et al. (1973), are:

i) the above salt intrusion, since $E < 0$, gives a pressure increase in the aquifer and in turn a consolidation of the aquitards. This can lead to surface subsidence since in this basin the coefficient of compressibility is $a_v \approx 3.5\text{x}10^{-6}$ in SI. Bonafede (1991) discussed similar phenomena in connection with the bradyseismic crisis in the Campi Flegrei.

ii ) the increase of Na C1 concentration in the Oxnard aquifer would tend to drive Na Cl into and through the adjacent aquitard by diffusion or advection. Thus Na Cl would contaminate adjacent aquitard and the contiguous Mugu aquifer.

To discuss the above points, we state that the presence of the marine salt in the upper layer, with a density as large as $\rho^* \sim 1.6$, gives very large $\boldsymbol{R} \sim 400$. Our model thus can allow a tentative forecast, since the geological properties of the basin allow the presence of transients of salt density. As a consequence a rather sharp and quick movement of marine salt can probably take place in these basins.

## 11. A "theoretical experiment" about osmosis in sandstones

The approach for dealing with rocks where osmosis has not yet been measured, but could potentially occur following Alexander (1990), Neuzil and Provost (2009) and Hart (2012), can focus on effects eventually revealing its presence. It could thus be possible to consider $\omega_0$ and Θ as very small quantities, but still allowing our theory to be applied. Here we therefore assume both $\omega_0$ and Θ to be very small, namely to have $\beta$ times the common values for shales, where $\beta$ is a very small parameter, and analyze possible osmotic properties of sandstones.

### *St. Peter Sandstone*

The St. Peter Sandstone is an Ordovician formation. The formation spans north-south from Minnesota to Missouri and east-west from Illinois into Nebraska in the Midwest of United States. The data discussed here were obtained from the regional studies in Minnesota, USA (Kanivetsky, 1978; Kanivetsky and Walton, 1979; Freeze and Cherry, 1979). This sandstone consists of massive, fine-to-medium-size, well-rounded quartz grains, well sorted and friable. The mineralogy of the sandstone is of almost purely (99% in weight) Quartzite. The sandstone is saturated at 50-60 m depth (Kanivetsky, 1978; Kanivetsky and Walton, 1979).

**Table VI$_a$.** Estimated values of relevant parameters for St. Peter Sandstone, in SI.

| Parameters | Values in SI | Units |
|---|---|---|
| $\varphi$ Rock porosity | 0.27 | / |
| $k$ Intrinsic permeability | $5 \times 10^{-14}$ | $m^2$ |
| Θ Solute reflection coefficient | $0.2\,\beta$ | / |
| *D* Solute diffusion coefficient | 4x10$^{-7}$ | m$^2$/s |
| *Ms* Solute molar mass (Na Cl) | 0.1 | kg/mol |
| $\alpha$ Biot coefficient | 0.5 | |
| $\omega_0$ Swelling coefficient | 4x10$^4\beta$ | Pa |
| *K* Bulk modulus | 2 x 10$^7$ | Pa |
| *Ks* Bulk modulus of the solid matrix | 4x10$^7$ | Pa |
| $\bar{\rho}$ Estimated solute density | 0.25 /0.5 | kg/m$^3$ |
| $\bar{p}$ Estimated pore pressure | 3x 10$^5\beta$ | Pa |

In these rocks the value of the bulk modulus *K* has a large variability, since apparently $K \approx$ 1.8-111x10$^{-6}$ while $K < K_s \approx$ 4 x10$^7$ in SI. We therefore tentatively assume a value of $K \approx$ 2 x 10$^7$ but with a very large uncertainty, while for the Biot coefficient we assume $\alpha$ = 0.5 and take the average value $\bar{\rho}$ = 0.3.

**Table VI$_b$.** Coefficients of equations (1) for St. Peter Sandstone, in SI.

| Coefficient | Numerical value in SI |
|---|---|
| $E=-\dfrac{\omega_0}{\alpha^2+KV}\left(\dfrac{1}{\bar{\rho}_S}-\dfrac{1}{\bar{\rho}_D}\right)$ | -3x10$^5$ m$^2$/s |
| $F=-\left(\dfrac{k}{\eta}K+\dfrac{k\Theta K\bar{\rho}_f}{\eta\bar{\rho}_D}\right)$ $/(\alpha^2+VK)$ | -2x10$^{-10}$ m$^2$/s |
| $H=\dfrac{k\Theta\overline{\rho}_f R^*TK}{\eta M^S(\alpha^2+KV)}(\dfrac{1}{\overline{\rho}_D}+\dfrac{1}{\overline{\rho}_S})$ | 7x10$^{-5}$ m$^4$/s |
| $M= -\dfrac{k}{\phi\eta}[\dfrac{1}{\overline{\rho}_f}+\dfrac{\Theta}{\overline{\rho}_D}]\overline{\rho}_f$ | -5x10$^{-17}$ m$^2$/s Pa |
| $N= \dfrac{k}{\eta\phi}\Theta\dfrac{R^*T}{M^S}[\dfrac{1}{\overline{\rho}_D}+\dfrac{1}{\overline{\rho}_S}]\overline{\rho}_f$ | 10 $^{-11}$m$^4$/s$^3$ Pa |
| $S= -\dfrac{D}{\phi}$ | -10$^{-9}$ m$^2$/s |
| $U=\dfrac{M^S D\overline{\rho}_S}{R^*T\phi\ \overline{\rho}_f}$ | 7.5x10$^{-16}$ s |
| $V=\dfrac{\alpha-\varphi}{K_s}+\dfrac{\varphi}{K_f}-\dfrac{\omega_0}{K}\ \dfrac{M^S}{R^*T\,\overline{\rho}_D}$ | - 2.5x10$^{-9}$1/Pa |

All this finally gives $A = N - EM \approx 10^{-2}$ and $Z = EU\text{-}S \approx 1.5 \times 10^{-6}$ and consequently $\boldsymbol{R} \approx 10^5$. This is due to two opposite effects, the decrease of osmosis and the corresponding increase of permeability. Therefore, also for an initial external stress as large as $\rho\beta \approx 6$ one can have shock solitary waves. Our final result is that solitons could be possible also in St. Peter sandstone, but again for a very large external stress.

***Jordan Sandstone***

The late Cambrian Jordan Sandstone Formations composed of a white to yellow, quartzose, fine- to coarse-grained sandstone, varying from friable to well cemented. The Jordan Sandstone is one of the major sources of groundwater in the Midwest and its extension is similar to St. Peter Sandstone.

The data discussed here were obtained in Minnesota (Kanivetsky, 1978; Kanivetsky and Walton, 1979; Freeze and Cherry1979). The mineralogy of these rocks is dominated by quartz (90% in weight). The sandstone is saturated at 70-80 m depth with the values shown in Table $VI_a$.

**Table $VII_a$.** Estimated values for relevant parameters of Jordan Sandstone, in SI.

| Parameters | Values in SI | Units |
|---|---|---|
| $\varphi$ Rock porosity | 0.31 | / |
| $k$ Intrinsic permeability | $5x10^{-12}$ | $m^2$ |
| $\Theta$ Solute reflection coefficient | $0.2\,\beta$ | / |
| $D$ Solute diffusion coefficient | $2x10^{-9}$ | $m^2/s$ |
| $M^s$ Solute molar mass (Na Cl) | 0.06 | kg/mol |
| $\alpha$ Biot coefficient | 0.5 | / |
| $\omega_0$ Swelling coefficient | $5x10^4\beta$ | Pa |
| $K$ Bulk modulus | $2x10^7$ | Pa |
| $Ks$ Bulk modulus of the solid matrix | $3.8x10^7$ | Pa |
| $\bar{\rho}$ Estimated solute density | 0.5 | $kg/m^3$ |
| $\bar{p}$ Estimated pore pressure | $4x10^5\beta$ | Pa |

From these quantities we compute

**Table VII$_b$.** Coefficients of equations (1) for Jordan Sandstone, in SI.

| Coefficient | Numerical value in SI |
|---|---|
| $E = -\frac{\omega_0}{\alpha^2 + KV}(\frac{1}{\bar{\rho}_S} - \frac{1}{\bar{\rho}_D})$ | -3x10$^5$ m$^2$/s |
| $F = -(\frac{k}{\eta}K + \frac{k\Theta K\bar{\rho}_f}{\eta\bar{\rho}_D})\ /(\alpha^2+VK)$ | -2x10$^{-10}$m$^2$/s |
| $H = \frac{k\Theta\bar{\rho}_f R*TK}{\eta M^S(\alpha^2 + KV)}(\frac{1}{\bar{\rho}_D} + \frac{1}{\bar{\rho}_S})$ | 7x10$^{-5}$m$^4$/s |
| $M = -\frac{k}{\phi\eta}[\frac{1}{\bar{\rho}_f} + \frac{\Theta}{\bar{\rho}_D}]\bar{\rho}_f$ | -5x10$^{-17}$ m$^2$/s Pa |
| $N = \frac{k}{\eta\phi}\Theta\frac{R*T}{M^S}[\frac{1}{\bar{\rho}_D} + \frac{1}{\bar{\rho}_S}]\bar{\rho}_f$ | 10 $^{-11}$m$^4$/s$^3$ Pa |
| $S = -\frac{D}{\phi}$ | -10$^{-9}$m$^2$/s |
| $U = \frac{M^S D\bar{\rho}_S}{R*T\phi\ \ \bar{\rho}_f}$ | 7.5x10$^{-16}$ s |
| $V = \frac{\alpha-\varphi}{K_s} + \frac{\varphi}{K_f} - \frac{\omega_0}{K}\frac{M^S}{R*T\bar{\rho}_D}$ | - 2.5x10$^{-9}$1/Pa |

From Table VII$_b$ we obtain $A = N - EM \approx 0.5\,\beta$ and $Z = EU - S \approx 6$ x10$^{-9}$ while $\boldsymbol{R} \approx \beta\rho*10^8$ as initial external stress. Consequently, our “theoretical experiment” confirms that in these sandstones one can find shock solitary waves. It is however rather surprising that solitons can also be present for such a particularly small $\beta$.

In summary, these findings support the original idea of a small presence of osmosis also in higher permeability rocks following Alexander, 1990; Neuzil, 2009; Neuzil and Provost, 2000; among others. Because this is the "theoretical experiment" we feel that field and laboratory analyses are needed to support our theory for higher permeability rocks.

## 12. Discussion and conclusions.

This paper analyses the action of a strong external stress on the dynamics of non-linear transients of $\rho$ and $p$ in geologic porous media. In particular, we focus on advection in low permeability rocks. As a result we obtain a non-linear model (i.e. two equations in 1-D) of the evolution of transients of combined $\rho$ and $p$. Their solutions unveil the presence of quick large shocks, known in physics as Burgers solitons (Whitham, 1974), which are ruled by the Reynolds number $\boldsymbol{R} = 2A\rho^*/Z$. A strong initial stress is needed to obtain $\boldsymbol{R} > 8\text{-}10$, the condition for having such non linear solitons. A totally novel point is that we are not only consider advection, but also non-linear effects in the Hooke law correlating strain, pore pressure and pollutant density.

Our model thus identifies a very simple analytical relation for estimating an eventual soliton transports in porous media, much quicker and sharper than those predicted by linear models, perturbation theories or scale analyses. When we compute $\boldsymbol{R} = 2A\rho^*/Z$ for various shales and clays, we found a surprising large $\boldsymbol{R}$ variability. This was *a priori* unexpected in rocks that are usually considered to be similar. But we also observed that $\boldsymbol{R}$ is roughly proportional to the ratio of permeability over diffusivity $k/D$.

Checking such estimated ***R*** for the rocks analyzed in this paper, we found that this rather unexpected variability was in reality due to the intrinsic *k/D* rock characteristics, i.e., as a result of their origin, depositional environment and subsequent evolution of porous media and are not related to some model characteristics. Thus it can easily evaluated from this ratio *k/D* whether or not solitons can be frequent in a given rock. This information could explain a number of geological processes, including fast hydrocarbon migration (Appold and Nunn, 2002; Joshi et al., 2012), development of epikarst environments (Dragila et al., 2016). Potential applications to the protection of water resources (salinization and pollution), nuclear waste disposal (see also Kim et al., 2011; Gonçalvès et al., 2012) and borehole drilling (see also Zhang, 2011; Zeynali, 2012) are foreseen.

About the other non-linear effect, i.e. the non linear pressure in the generalized Hooke law, we obtain that in the very early moments the pressure is increased by $-\frac{\eta^{*} E^{2} N}{16\, A}\,\frac{x^{4}}{t^{2}} \approx -\,10^{8}\,\eta^{*}\frac{x^{4}}{t^{2}} > 0$. The novel term is proportional to non linear pressure parameter $\eta^{*} < 0$. To our knowledge, this effect has never been discussed and, rather surprisingly the various behaviors of different rocks are mainly due to this parameter $\eta^{*}$ since the numerical coefficient of rocks are rather similar.

To analyse if osmotic phenomena can be present in geological porous media more widely than previously thought, we finally discuss a “theoretical experiment”. We assume for sandstones that the swelling and solute diffusion coefficients (i.e. the main quantities related to osmosis) are as those for clay but multiplied by a small, or also very small, heuristic parameter $\beta$.

In this way we obtain that one can also have rather large Reynolds numbers as for Jordan and St. Peter sandstone, thus supporting the formation of solitons also in higher permeability rocks.

Although osmotic constants and properties of geologic porous media are poorly understood, although osmotic processes may occur in geological processes, such as fluid migration (Magara, 1974) and be important in solution of many practical problems, such as nuclear wasete disposal (Kim et al., 2011), borehole drilling (Schlemmer et al., 2003) and shale gas production (Engelder et al., 2014) it is well known fact that osmotic constants and properties of geologic prous media are poorly understood. Therefore, our analysis is an attempt to estimate some relevant characteristics of hydrogeologic system. We, moreover, suggest the need for further field and laboratory studies in order to improve solitary waves theory that employs non linear models.

**Acknowledgements**

The authors would like to thank Prof. Carminati for warm help. Prof C. Doglioni and Dr. M. Sommacal for providing essential information.

**Figure captions**

Fig 1. Solutions of Burgers equation, with ***R*** = 1, curve 1; ***R*** = 3, curve 2; ***R*** = 5, curve 3; ***R*** = 7, curve 4; ***R*** = 9, curve 5; ***R*** = 25, curve 6 (suggested also by Merlani et al., 2011).

Fig 2. Schematic variation of $\rho$ with distance at different times (suggested also by Merlani et al., 2011).

Fig 3. Schematic variation of function $u$ with distance at various times. The symbol $u$ is in reality just $x/t$. It is useful to compute the non linear velocity $u_B = (Ex)/(At)$, while the Darcy fluid velocity $u_D = (k\, Ex)/(\varphi\eta\, At)$ is much smaller (suggested also by Merlani et al., 2011).

Fig 4. Two typical stress-strain curves: panel a) ductile behavior and panel b) brittle behavior, for materials under uniaxial tension are schematically shown (suggested also by Gross and Seeling, 2006).

**Fig 1**

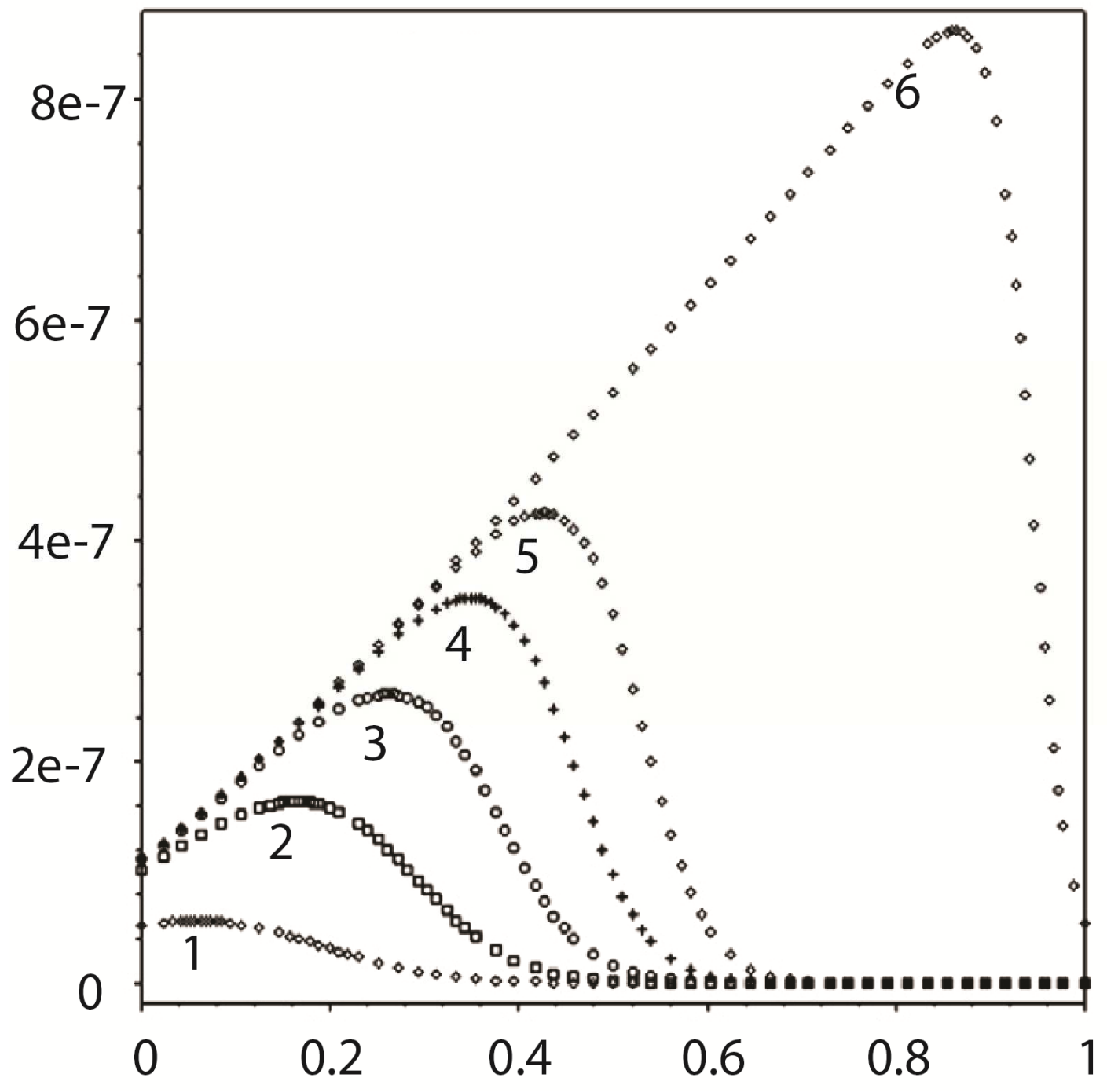

**Fig 2**

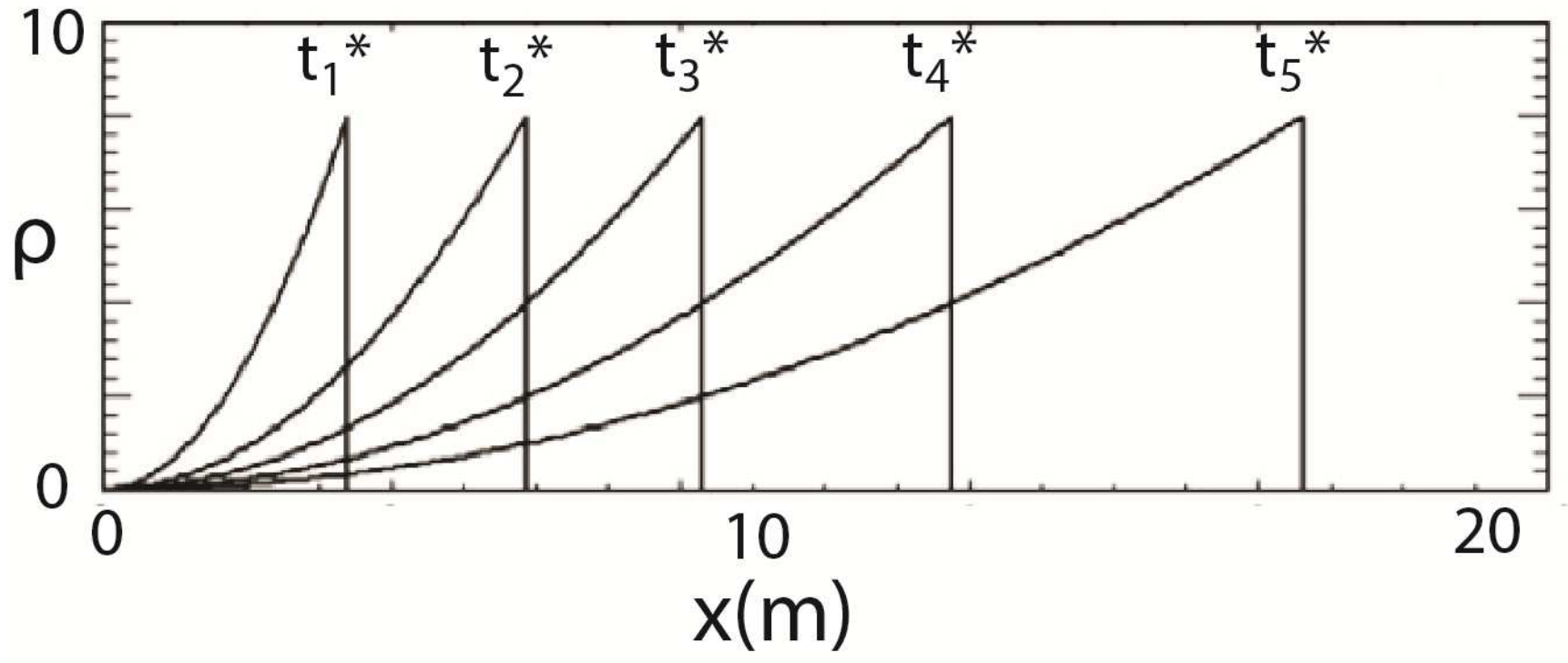

**Fig. 3**

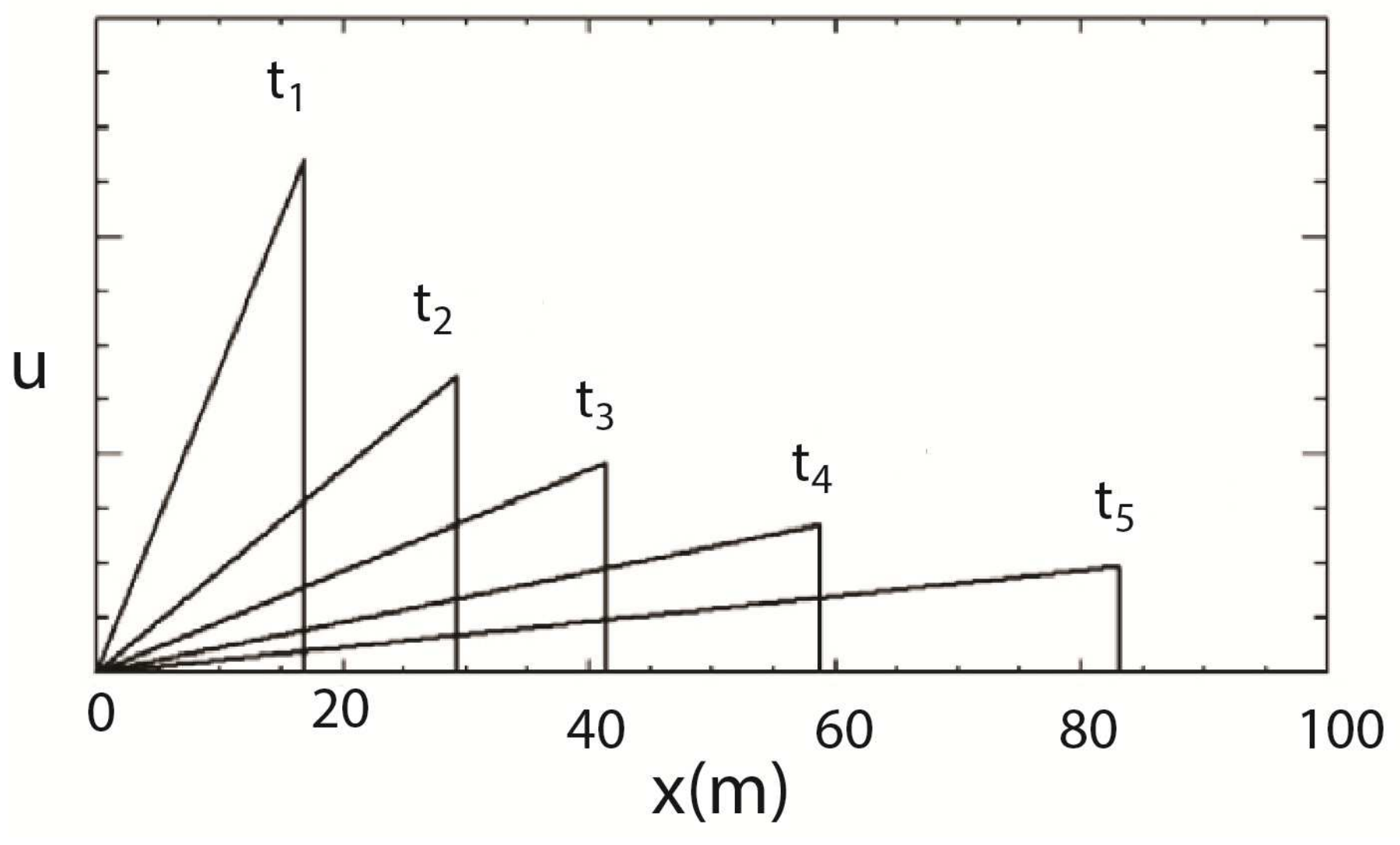

**Fig 4**

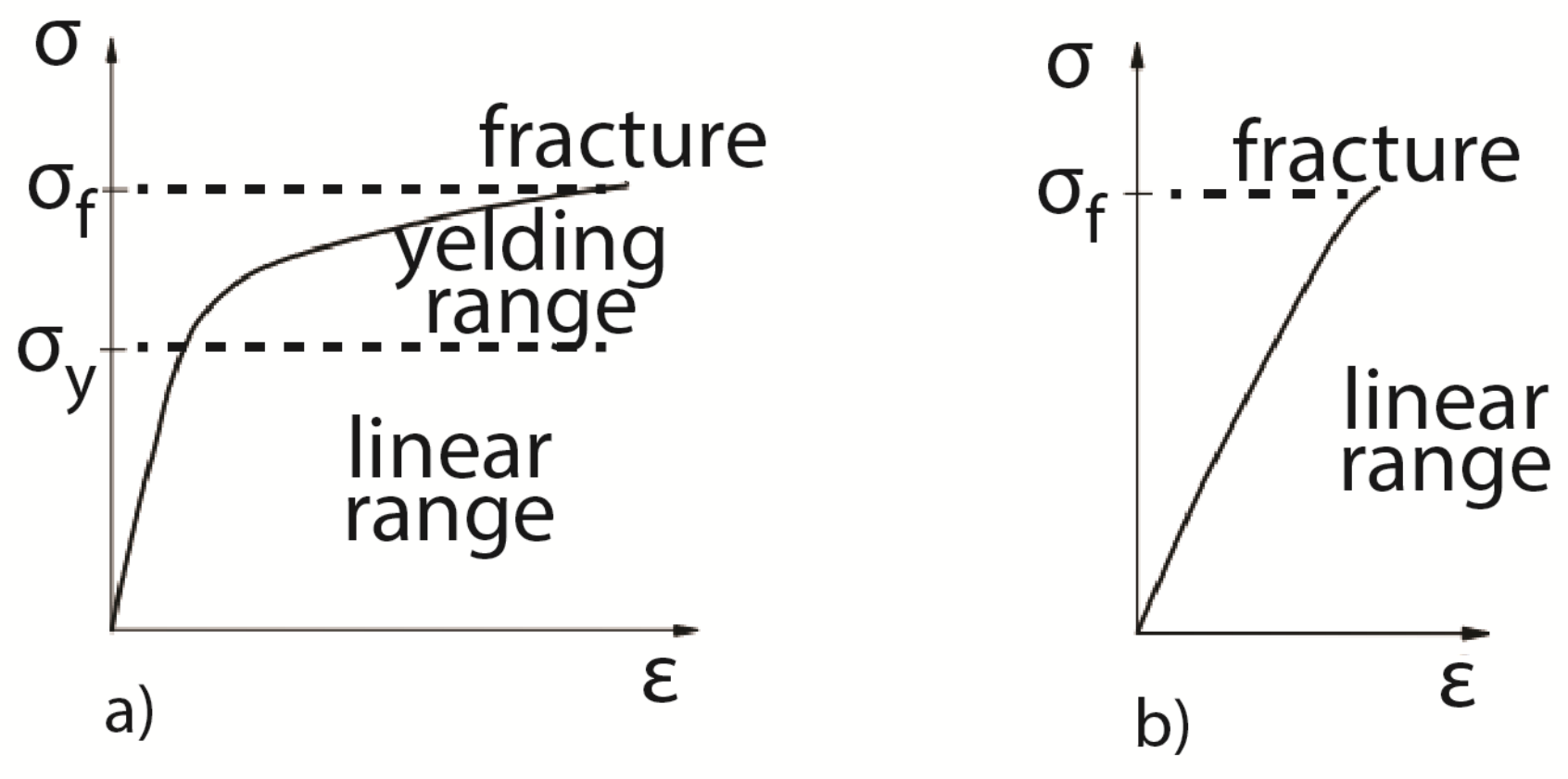

**Appendix A. The Ghassemi and Diek (2003) and Caserta et al. (2010) model**

***Transport equations***

The conceptual model discussed in this study is based on extended versions of the equations of poroelasticity and Darçy transport law, which are derived using a non-equilibrium thermodynamics approach for dilute solutions of salt in a fluid. Their linearized transport equations for pore fluid pressure $p$ and solute contribution to total fluid density $\rho$ are (GD03)

$$\frac{\partial \zeta}{\partial t}\bar{\rho}_f + \nabla \cdot J_f = 0 \qquad \phi\frac{\partial \rho}{\partial t} + \frac{1}{\bar{\rho}_f}\left(J_f \cdot \nabla\rho\right) + \nabla \cdot J_S = 0 \qquad \text{(A1)}$$

where the overbar means an averaged quantity. In more detail here $\rho_f$ is the fluid density, $J_f$ the fluid flux, $J_s$ the solute flux, $\varphi$ the rock porosity, $u$ the fluid velocity, $\zeta$ the rock fluid content per unit rock referential volume. The relation among $\zeta$, the pore volume fraction $v$ and $\varphi$ are discussed in the following.

The phenomenological "forces" driving an ideal binary solution are gradients of pore pressure $p$ and chemical potential $\mu^S$ and $\mu^D$, namely

$$J_f = -\frac{\bar{\rho}_f k}{\eta}\left[\nabla p - \bar{\rho}_f \Theta \nabla\left(\mu^S - \mu^D\right)\right] \qquad J_S = -\frac{M^S D \bar{\rho}_S \rho_D}{R^* T \bar{\rho}_f}\nabla\left(\mu^S - \mu^D\right) \qquad \text{(A2)}$$

where $k$ is the intrinsic permeability, $\eta$ the fluid viscosity, $\Theta$ the standard solute reflection coefficient ($0 < \Theta < 1$, GD03), $\mu^S$ is the solute chemical potential , $\mu^D$ is the diluent chemical potential, $D$ the solute diffusion coefficient, $R^*$ is the universal gas constant, $T$ is the absolute temperature, $\rho_D$ is the solvent density, $M^S$ the molar mass of the solute, often Na Cl.

Equations (A1) and (A2) generalize the Darçy law by recognizing that gradients in the chemical composition of the pore fluid can be a force driving molecular or grain filtration. Such force can discriminate (or favors) transport of a specific component as an osmotic forcing through semipermeable membranes. Expressions (A1) and (A2) show that hydraulic gradients give hydraulic diffusion while the chemical potential is related to chemical pressure diffusion (osmosis). One has moreover to remark that in (A1) and (A2) the chemical potentials are related to osmosis while the Laplacian $\Delta\, p$ and $\Delta(\mu^s - \mu^D)$ schematize the diffusion of pore fluid and solute density.

***Osmotic effects***

We focus first our attention on low permeability clay and shales: the purpose of this study is to emphasize how the concomitant presence of solute and solvent in permeable rocks can remarkably affect the fluid transient propagation. About osmosis for a rather dilute ideal solution, GD03 assume a linearized solute chemical potential (Table $I_a$)

$$\mu^S = \frac{R^* T}{M^S} \ln \frac{\rho}{\bar{\rho}_f} \approx \frac{R^* T}{M^S}\left(1 + \frac{\rho}{\bar{\rho}_f}\right) \qquad \text{(A3)}$$

while from the Gibbs-Duhem relation for the diluent they show that

$$\mu^D = \frac{p}{\bar{\rho}^D} - \mu^S \frac{\rho}{\bar{\rho}^D} \approx \frac{p}{\bar{\rho}^D} - \frac{R^* T \rho}{M^S \bar{\rho}^D} \qquad \text{(A4)}$$

***The stress, strain and pore volume fraction equations***

The dynamical effect of pressure/solute density jumps on the porous rocks can affect remarkably the matrix permeability, porosity, diffusion and so on, as described on phenomenological grounds by Neuzil and Provost (2009) among others. Therefore we analyze the general constitutive equation for a saturated fluid with a contaminant substance in an isotropic porous rock. The linearized equation for stress $\sigma_{ij}$ is

$$\sigma_{ij} = 2G\epsilon_{ij} + \left( K - \frac{2G}{3} \right) \sum_k \epsilon_{kk}\delta_{ij} - \alpha p \delta_{ij} + \left( \omega^D \mu^D + \omega^S \mu^S \right) \delta_{ij} \qquad \text{(A5)}$$

Here $\sigma_{ij}$ is the total stress tensor, $\varepsilon_{ij}$ is the strain tensor, $G$ and $K$ are the shear and bulk modulus, respectively; $\omega^k$ is a chemical-mechanical coupling coefficient of the $k^{th}$ substance (only a solute and a diluent in our case), the poroelastic (Biot) coefficient is $\alpha$ =1 - $K/K_s$, where $K_s$ is the bulk modulus of the solid matrix.

In equation (5) appear the chemical-mechanica1 coupling coefficients $\omega^D$ and $\omega^S$ for just one contaminant, that following GD03 are

$$\omega^S = \omega^D = \omega_0 \frac{M^S}{RT} \qquad \text{(A6)}$$

where $\omega_0$ is the swelling coefficient (Bader and Kooi, 2005). Therefore, from (A3), (A4) and (A5) one can express

$$\omega^S \mu^S + \omega^D \mu^D \approx \omega_0 \;\; \left(\frac{1}{\bar{\rho}} - \frac{1}{\bar{\rho}_D}\right)\rho + \frac{\omega_0\, M^S}{\bar{\rho}_D R * T}\, p \qquad \text{(A7)}$$

We remark how different approaches are present in the literature, that give apparently different formulations but each approach finally approximates both $J_s$ and $J_f$ as linear functions of $p$ and $\rho$. Thus these different formulations can change the numerical values of the rock parameters present in (A1)....(A6) but the structure of the equations remains unchanged.

To connect (A3), (A4) or (A8) to equations (A1) and (A2) in order to obtain the explicit evolution equations, GD03 remark how the variation of rock fluid content $\zeta$ can be expressed as

$$\frac{\partial \zeta}{\partial t} = \frac{\rho \partial v}{\overline{\rho_f} \partial t} + \frac{v \partial p}{K_f \partial t} \approx \frac{\partial v}{\partial t} + \frac{\phi \partial p}{K_f \partial t} \qquad \text{(A8)}$$

where $K_f$ is the fluid bulk modulus.

The equi1ibrium equation in a full 3-dimensional formulation represents the momentum balance disregarding as usual the inertia and body forces

$$\Sigma\, \partial\, \sigma_{ij} / \partial x_i = 0 \qquad \text{(A9)}$$

An important class of problems entails one-dimensional deformation as, for example, a planar wall or half–space. In such a simple case one has that for symmetry the stress is constant (Mc Tigue, 1986). Consequently the 1-D assumption finally implies (MSV)

$$K\, \Sigma_k\, \varepsilon_{kk} - \alpha\, p \;+\; \left(\frac{1}{\overline{\rho}} - \frac{1}{\overline{\rho}_D}\right)\rho \;+ \frac{\omega_0\, M^S}{\overline{\rho}_D R^* T}\, p = 0 \qquad \text{(A10)}$$

that can also hold locally for punctual (spherical coordinates) and cylindrical (cylindrical coordinates) sources.

From these relations our main equations (1) are obtained.

## Appendix B .The structure of the fronts

We here analyze some properties of the Burgers-like equation, this is not a formal mathematical demonstration but just an intuitive but exact sketch. Consider in general the equation

$$\frac{\partial T}{\partial t} = D \frac{\partial^2 T}{\partial z^2} + M(T) \left(\frac{\partial T}{\partial z}\right)^2 \qquad \text{(B1)}$$

with $D$ constant and define $Q = \partial\, T / \partial\, z$. By $z$-deriving (B1) we have

$$\frac{\partial Q(T)}{\partial t} - D\frac{\partial^2 Q(T)}{\partial z^2} - M(T)\frac{\partial Q(T)^2}{\partial z} - \frac{\partial M}{\partial z}Q^2 - 2M\frac{\partial Q^2(T)}{\partial z} = 0 \qquad \text{(B2)}$$

Assuming that $T= T_0$ is constant in a small region around $z \approx a$ and $T = T_0+ T_I$ is again constant around $z \approx b$ we thus have that $Q(a) = Q(b) = 0$ in the above regions. In turn another $z$-derivative of equation (B2) gives that in small regions around $z = a$ and $z = b$ one has $\frac{\partial^2 Q}{\partial z^2} = M^2\frac{\partial (Q^2)}{\partial z} = \frac{\partial Q}{\partial z} = 0$. Once integrated between $a$ and $b$ the relation (B2) therefore gives

$$\int_a^b \frac{\partial Q}{\partial t}dz = \frac{\partial}{\partial t}\int_a^b Q\, dz = \frac{\partial}{\partial t}[T(b) - T(a)] = 0 \qquad \text{(B3)}$$

that implies that $T(b) - T(a) = T_0+ T_I - T_0 = T_I = const.$ For $T_0$ fixed this implies that $T(b) = T_I$.

If the solution of (B1) is growing like a polynomial $z, z^2,\ldots$ in the $a$ - $b$ interval and we fix that $a = 0$ and $b = z_B$, to satisfy the equation (B 3) we must assume $T(z_B,t) = T_0+ T_I$ and just a flat $T(z,t) = T_0$ for $z > z_B(t)$, in particular if $t \rightarrow 0$ and $z \rightarrow 0$.